\pdfoutput=1

\documentclass[sigconf,nonacm]{acmart}

\usepackage{booktabs}
\usepackage{tabularx}
\newcolumntype{L}[1]{>{\RaggedRight\arraybackslash}p{#1}}

\usepackage{array}
\usepackage{ragged2e}

\AtBeginDocument{%
  }

\author{Atieh Taheri}
\affiliation{%
 \institution{Carnegie Mellon University}
 \city{Pittsburgh}
 \state{Pennsylvania}
 \country{USA}}
\email{ataheri@cmu.edu}
\orcid{0009-0008-8815-7809}

\author{Mahya Tazike}
\affiliation{%
 \institution{Indiana University Indianapolis}
 \city{Indianapolis}
 \state{Indiana}
 \country{USA}}
\email{mtazike@iu.edu}
\orcid{0000-0001-6757-666X}

\author{Patrick Carrington}
\affiliation{%
 \institution{Carnegie Mellon University}
  \city{Pittsburgh}
  \state{Pennsylvania}
  \country{USA}
}
\email{pcarring@cmu.edu}
\orcid{0000-0001-8923-0803}
  
\author{Jeffrey P. Bigham}
\affiliation{%
  \institution{Carnegie Mellon University}
  \city{Pittsburgh}
  \state{Pennsylvania}
  \country{USA}
}
\email{jbigham@cmu.edu}
\orcid{0000-0002-2072-0625}

\begin{document}


\title{When Disability Disclosure Travels: Memory, Privacy, and Contextual Integrity in Conversational AI}

\renewcommand{\shortauthors}{Taheri et al.}

\begin{abstract}
  Conversational AI assistants remember what people tell them, and for disabled people, that often includes disability. We interviewed 12 adults with disabilities in the United States who use LLM-based assistants such as ChatGPT, Claude, and Gemini about when, how, and why they disclose disability to these systems and how this compares with disclosing to people. Using contextual integrity as an analytic lens, we found that participants disclosed by need rather than by name, translating disability into task-scoped instructions; that the same disclosure was judged against two recipients, a non-judging interlocutor and a data-holding company, producing opposite norms; and that memory features relieved the burden of repeated disclosure while letting disability information drift into contexts where it did not belong. Participants did extensive boundary work to restore context and wanted control over scope, provenance, retention, and access rather than per-utterance toggles. We discuss implications for the design of conversational AI assistants.
\end{abstract}





\maketitle

\section{Introduction}
For many disabled people, getting useful help from a conversational AI assistant begins with telling it something about disability. A blind user asks for a description ``the way you'd tell a blind person.'' A wheelchair user asks for date ideas that are wheelchair accessible. A person with an anxiety disorder asks for feedback that is constructive rather than critical. These are ordinary, practical instructions, but each one is also a disclosure, and disclosure of disability has never been a neutral act. Decades of scholarship describe the calculus disabled people perform before telling an employer, a classmate, a doctor, or a stranger about an impairment: weighing access against stigma, deciding how much to say and in what words, and repeating the decision every time the audience changes~\cite{goffman1963stigma,lingsom2008invisible,samuels2003body,evans2019trial}.

Conversational assistants change the terms of that calculus in two ways. First, the audience is strange. An assistant presents itself as an attentive, patient, non-judging interlocutor, and people do tell machines things they would not tell people~\cite{lucas2014computer,ho2018chatbot}; yet everything said to it is also received, stored, and potentially reused by a company, whose retention practices and secondary uses are opaque~\cite{zhang2024fairgame,lee2024taxonomy,mireshghallah2024trustnobot}. Second, the assistant now remembers. Cross-session ``memory'' features in ChatGPT, Gemini, and Claude are a form of personalization that carries what a user said in one conversation into later ones. For disabled users this can be exactly what is needed: an assistant that has been told once that the user reads with a screen reader, or cannot climb stairs, can format every later answer accordingly and suggest only places that are reachable, without being reminded each time. The same mechanism lets information travel. A diagnosis mentioned while asking about a medication can reappear while planning dinner, or be applied to a question asked on someone else's behalf. Memory is thus at once a relief and a new way for disability information to end up somewhere it was never meant to go.

HCI has begun to document how disabled people use generative AI: as a screen reader of last resort, a writing aid, a visual interpreter, a tutor, and sometimes a confidant~\cite{glazko2023autoethnographic,adnin2024king,mullen2024diary,valencia2023less,jang2024onlything}. Parallel work shows that these systems reproduce ableist assumptions~\cite{gadiraju2023offensive,hutchinson2020social,phutane2025cold,mack2024wheelchair} and that memory and personalization raise privacy concerns for users generally~\cite{chen2026relational,jo2024memory,zhang2025ragmemory}. What is missing is an account of how disabled people reason about \emph{disclosing disability} to assistants that remember: what they say, in what form, to whom they imagine they are saying it, and what happens when the assistant applies, fails to apply, or over-applies what it was told.

We address this gap through semi-structured interviews with 12 disabled adults in the United States who use LLM-based assistants at least twice a week, or who once did and have since cut back. Participants included people who are blind or have low vision, people with mobility and physical disabilities, people who are d/Deaf or hard of hearing, and people with chronic illness, mental health, and attention-related disabilities; two thirds reported more than one. The interviews were conducted by a disabled researcher who disclosed her own disability during sessions, which shaped the conversations in ways we discuss. We analyze the interviews through Nissenbaum's theory of contextual integrity~\cite{nissenbaum2004privacy,nissenbaum2010privacy}, which treats privacy as the appropriateness of information flows relative to the norms of the context in which the information was shared. Contextual integrity is well suited to disability disclosure because disclosure is rarely a question of secrecy; it is a question of whether a given piece of information, shared for a given purpose, should travel to a given recipient under given conditions. We asked:

\begin{itemize}
  \item \textbf{RQ1.} What do disabled adults disclose about disability to conversational AI assistants, in what form, and how do they decide?
  \item \textbf{RQ2.} How do these decisions compare with disclosing to people, and what norms do participants apply to the assistant as a recipient?
  \item \textbf{RQ3.} How do memory and personalization affect the appropriateness of disability-related information flows, and what forms of control do participants want?
\end{itemize}

Our findings make three contributions. Empirically, we provide a cross-disability account of disclosure to conversational AI, showing that participants disclosed \emph{by need rather than by name}, translating disability into task-scoped instructions and reserving diagnoses for medical tasks or for cases where they expected the label to carry knowledge the assistant should already have. Analytically, we show that a single disclosure to an assistant is evaluated against \emph{two recipients} at once, the agent and the company behind it, and that whether participants disclosed more or less to AI than to people differed according to which recipient they foregrounded; and that memory changes a disclosure's transmission principle from ephemeral and scoped to persistent and global, which participants experienced as both relief from repetition and a drift of disability information into inappropriate contexts, countered through boundary work. For design, we translate the findings into implications at three layers of an assistant, distinguishing those supported by several participants from those we offer tentatively; together they favor controls over scope, access, retention, and provenance rather than per-utterance toggles, which four of the eight participants who discussed them declined.

\section{Related Work}

\subsection{Disability Disclosure as Information Management}

Goffman's account of stigma frames disclosure as the management of discreditable information: people whose stigmatized attribute is not immediately apparent must decide, again and again, whether to tell, and to whom, how, when, and where~\cite{goffman1963stigma}. Disability scholars have elaborated this into a rich literature on concealment and disclosure. Lingsom describes the dilemmas of people with invisible impairments, who cannot rely on others to already know and who therefore carry the burden of deciding whether and how to inform them~\cite{lingsom2008invisible}. Samuels argues that the ``coming out'' metaphor borrowed from queer discourse fits disability only partially, because disability disclosure is repeated, contextual, and often compelled by the need for access rather than chosen as an act of identity~\cite{samuels2003body}. Evans distinguishes forms of impairment disclosure and shows how disclosure, and the reactions it provokes, shape disability identity over time~\cite{evans2019trial}.

Workplace studies show that disclosure decisions turn on anticipated consequences: employees weigh accommodation against discrimination, and disclose more where organizational climate signals safety~\cite{vonschrader2014perspectives,santuzzi2014invisible,brohan2012systematic}. Online, disabled people disclose strategically, choosing platforms, audiences, and phrasings~\cite{furr2016strategic}. Even assistive technology itself is a form of disclosure: Shinohara and Wobbrock show that visible assistive devices invite misperception and that users manage the social accessibility of their tools alongside their functional accessibility~\cite{shinohara2011shadow,shinohara2016selfconscious}. Two threads of this literature matter for what follows. Disclosure is frequently \emph{instrumental}, made to obtain access rather than to express identity; and it is \emph{recurrent}, re-performed for each new audience. Both threads reappear, transformed, when the audience is an assistant that neither asks nor forgets.

\subsection{Privacy, Accessibility, and the Tradeoffs Disabled People Are Asked to Make}

Privacy research with disabled people has repeatedly found an accessibility--privacy tradeoff. People with visual impairments report concerns about who can see their screens and their data, and adopt workarounds to manage them~\cite{ahmed2015privacy}. Camera-based assistive applications ask blind users to share images they cannot inspect, exposing them and bystanders to risks they can only partly assess~\cite{akter2020uncomfortable,akter2022shared,stangl2020visual,stangl2022privacy,alharbi2022obfuscation}. Adaptive assistive technologies collect behavioral data that is itself sensitive; Hamidi et al.\ show that users of adaptive pointing tools distinguish sharply between recipients and purposes when deciding who should have access~\cite{hamidi2018pointing}. Voice assistants were adopted by disabled users in part ``by accident,'' as an accessibility gain, even as the same devices raised always-listening concerns for users generally~\cite{pradhan2018accessibility,lau2018alexa}.

Sannon and Forte's review of privacy research with marginalized groups argues that privacy costs fall unevenly, that marginalized users often lack the option to opt out, and that research should attend to the strategies people develop rather than only to their concerns~\cite{sannon2022privacy}. Recent work extends this to generative AI: blind and low-vision people increasingly use multimodal models for tasks they previously relied on family for, including managing private visual content, and express a preference for on-device processing and zero retention~\cite{sharma2025beforemom}. In a 2026 American Foundation for the Blind survey, 60\% of respondents said they would prefer an AI reader over a human for sensitive documents if the tool neither saved nor shared the images, but only 17\% of disabled respondents would if the images were shared with a technology company~\cite{afb2026innovation}. Our study takes up the strategies these authors call for, in the specific case of disability information shared with assistants that remember.

\subsection{Disabled People and Generative AI}

Disabled users were early and intensive adopters of large language models. Autoethnographic and interview studies document blind people using ChatGPT and image-describing models as ``the king of knowledge''~\cite{adnin2024king}, neurodivergent ``power users'' building elaborate prompting routines and adapting AI tools to everyday tasks~\cite{glazko2025neurodivergent,xue2025characterizing,tazike2026helps}, autistic workers treating an LLM as ``the only thing I can trust'' for communication help~\cite{jang2024onlything}, users with disabilities treating chatbots ``kind of like a diary''~\cite{mullen2024diary}, and AAC users finding that language models can both enhance and impede communication~\cite{valencia2023less,goodman2022lampost}. Glazko et al.\ catalog the utility and the failures of generative AI for accessibility across tasks~\cite{glazko2023autoethnographic}, and studies of visual interpretation apps and live video assistance show high reliance alongside frequent errors and inappropriate assumptions about what a blind user can see~\cite{gonzalezpenuela2025mllm,chang2025probing,alharbi2024misfitting}.

The same models carry ableist assumptions. Language models associate disability with negative sentiment and toxicity~\cite{hutchinson2020social,venkit2022implicit,venkit2023automated}; disabled participants in Gadiraju et al.'s focus groups found LLM outputs subtly offensive and reductive~\cite{gadiraju2023offensive}; GPT-based resume screening penalizes disability-related credentials~\cite{glazko2024resume}; text-to-image models ``only care to show us the wheelchair''~\cite{mack2024wheelchair}; and models explain ableism in ways disabled people find ``cold, calculated, and condescending''~\cite{phutane2025cold}. These harms are traced to the disability models embedded in AI design~\cite{newmangriffis2023definition,whittaker2019disability,trewin2019considerations,bennett2020fairness} and to training procedures that reward agreement over accuracy~\cite{sharma2024sycophancy}. Disability studies scholarship in HCI asks that such systems be examined from lived experience, with attention to interdependence and access as ongoing work~\cite{mankoff2010disability,hofmann2020living,bennett2020care,tang2026cripping, taheri2025designing}. Our study sits at the seam between these two literatures: what disabled users tell assistants, and what assistants do with it.

\subsection{Disclosure, Memory, and Contextual Integrity in LLM Assistants}

Contextual integrity holds that information flows carry context-relative norms, specified by the type of information, the sender, recipient, and subject, and the transmission principle under which the information moves; a flow violates privacy when it departs from the entrenched norms of the context in which the information was shared~\cite{nissenbaum2004privacy,nissenbaum2010privacy}. The framework has been operationalized to elicit norms for smart home devices~\cite{apthorpe2018discovering} and, in the last three years, to evaluate and constrain language models: benchmarks test whether models keep secrets across contexts~\cite{mireshghallah2024secret,shao2024privacylens,li2025privacibench}, including persistent-memory-specific violations in CIMemories~\cite{mireshghallah2026cimemories}, and agent architectures use contextual integrity to minimize what an assistant reveals to third parties~\cite{bagdasarian2024airgapagent,ghalebikesabi2025privacy}.
Tran et al.\ used contextual integrity vignettes with 300 ChatGPT users and found that procedural safeguards such as anonymization weighed more in appropriateness judgments than the trustworthiness of the recipient~\cite{tran2025understanding}.
Shvartzshnaider and Duddu argue that much of this work applies the theory thinly, treating it as a checklist rather than as an account of norms grounded in the practices of actual communities~\cite{shvartzshnaider2025position}. We take the opposite route, using contextual integrity to interpret how one community reasons about flows. As we show, the norms participants articulated in depth were those governing the assistant as recipient; the norms they applied to the platform behind it were sparser and closer to the general institutional norms that prior privacy research describes.

The closely related notion of context collapse describes what happens when technologies merge audiences that were previously separate, so that a message composed for one is read by another~\cite{marwick2011tweet,vitak2012context}. Memory in conversational assistants produces a temporal version of the same problem. Studies of ChatGPT's memory feature find that users value the relational gains of being remembered while experiencing privacy strain and holding incomplete mental models of what is stored~\cite{chen2026relational, zhang2025ragmemory, yan2025remembering, malki2026usable}, and that long-term memory changes what people are willing to disclose~\cite{jo2024memory}. An analysis of 2,050 memory entries from 80 ChatGPT users found that 96\% had been created by the system rather than at the user's request, and that 35\% of users had health-related information saved in memory, accumulating into what the authors call an ``algorithmic self-portrait''~\cite{dash2026algorithmic}. Benchmarks now measure ``cross-domain leakage'' and ``memory-induced sycophancy,'' cases in which stored facts reappear where they do not belong or bend later answers~\cite{pulipaka2026persistbench,zhang2024ghost}. Zhang et al.\ show that users of LLM-based agents navigate disclosure by trading perceived benefits against poorly understood risks, and that the human-like interface itself encourages disclosure~\cite{zhang2024fairgame}; analyses of real conversation logs confirm that people share health details freely~\cite{mireshghallah2024trustnobot}. People disclose more to agents they believe are non-judging computers~\cite{lucas2014computer,ho2018chatbot,lee2020hearyou}, and an agent can be designed to extract personal information~\cite{zhan2025malicious}. Privacy in language models cannot be reduced to removing identifiers: models infer attributes from text~\cite{staab2024beyond,brown2022privacy,carlini2021extracting}, and the risks of aggregation and secondary use are distinct from exposure~\cite{lee2024taxonomy}.

This literature has studied disclosure to LLMs with general populations, and disabled people's use of LLMs without focusing on disclosure. Ours is, to our knowledge, the first study to ask disabled people across disability types how they decide what an assistant that remembers should know about their disability.

\section{Methods}

We conducted a qualitative interview study, approved by Carnegie Mellon University's IRB, with disabled adults who use LLM-based assistants. The study was exploratory and pre-specified contextual integrity as its theoretical frame; we had no hypotheses about which norms participants would apply.

\subsection{Recruitment and Participants}

We recruited participants through disability organizations and their mailing lists, disability-focused online community groups, and a national blindness organization's research participant request process. Recruitment materials described the study as being about everyday use of AI assistants, without naming disclosure, so as not to prime participants; the consent form and the interview introduction named disclosure as the focus. Interested individuals completed a short eligibility screener. Eligible participants were adults (18+) living in the United States who self-identified as having a disability and who either used an LLM-based assistant (for example, ChatGPT, Claude, Gemini, or Copilot) at least twice a week or had done so previously and had since stopped or substantially reduced their use; the second group was included as a contrast. We invited screener respondents in waves, prioritizing groups that were scarce among respondents (people who had reduced their use, people who are d/Deaf or hard of hearing, and people with multiple disabilities). All participants were offered accommodations at scheduling, including live captions, breaks, split sessions, screen-reader-compatible materials, and answering in writing in the meeting chat. Each participant received US\$20.

Twelve people participated (Table~\ref{tab:participants}). Six identified as women and six as men; ages ranged from 25--74. Nine identified as White, one as Black or African American, one as Native Hawaiian or Pacific Islander, and one as American Indian or Alaska Native, Hispanic or Latinx, and White. In an optional demographic questionnaire that mirrored the screener's disability categories, five participants selected blind or low vision, six selected mobility or physical disability, three selected d/Deaf or hard of hearing, five selected chronic illness, chronic pain, or fatigue, two selected attention-related disability, two selected mental health disability, and one each selected autism or another form of neurodivergence, cognitive disability, and speech or communication disability. Eight of the twelve selected more than one category. Seven had their disability since birth or early childhood, two for more than ten years, one for five to ten years, and two said it varied across their disabilities. Eleven used assistive technology regularly, from screen readers and braille displays to power wheelchairs, eye tracking, hearing aids, and AI-enabled smart glasses. Participants were, as a group, experienced with technology: several worked in software, web development, or assistive technology, and most had used LLM assistants for more than a year. Two participants (P6 and P9) had substantially reduced their use, and one (P10) described a cautious, deliberately limited use.

\begin{table*}[t]
\caption{Participants. Disability categories are those participants selected on the demographic questionnaire (multiple selections allowed); conditions are given only where participants named them. ``Memory'' summarizes each participant's account situation at the time of the interview.}
\label{tab:participants}
\small
\begin{tabularx}{\textwidth}{@{}l l X L{3.1cm} L{4.0cm}@{}}
\toprule
\textbf{ID} & \textbf{Age} & \textbf{Disability (Self-reported)} & \textbf{Assistants Used} & \textbf{Memory and Account Situation} \\
\midrule
P1 & 45--54 & d/Deaf or hard of hearing; mobility; speech or communication (since birth) & ChatGPT, Claude & Memory on; disability and access needs written into system instructions \\
P2 & 35--44 & Mobility (congenital muscular dystrophy; power wheelchair) & Claude, Gemini & Separate accounts for business, personal, and hobby use; profile in custom instructions \\
P3 & 25--34 & Mobility; chronic illness; attention-related (spinal muscular atrophy, Crohn's disease) & ChatGPT, Claude, Gemini, others; coding agent & Memory on; inspects memory through developer tools; syncs medical records \\
P4 & 25--34 & Mobility; chronic illness (spinal muscular atrophy and others) & ChatGPT & Memory on; told it about the wheelchair, not the diagnosis; never checked memory \\
P5 & 45--54 & Blind; chronic illness; attention-related; neurodivergence; mental health; cognitive & Siri, Alexa, ChatGPT & Free tier that ``always forgets''; uses a sibling's paid account \\
P6 & 45--54 & Chronic illness; mental health (multiple sclerosis and an MS-related anxiety disorder) & ChatGPT (web, no account), Copilot via clients & No account by choice; re-discloses in every session; reduced use \\
P7 & 25--34 & Mobility (C4 spinal cord injury; 5--10 years) & ChatGPT & Memory on; told it the injury level explicitly \\
P8 & 55--64 & Blind or low vision; chronic illness (congenital glaucoma; psoriatic arthritis) & ChatGPT, Be My Eyes, Seeing AI, Meta AI glasses, Alexa & Memory on; believes it inferred blindness from uploaded content \\
P9 & 35--44 & Blind (since birth) & ChatGPT, Claude, Meta AI glasses, Be My Eyes, Aira & Former paid subscriber; reduced use; prefers not to be remembered \\
P10 & 65--74 & Blind (since birth); hard of hearing (recent) & ChatGPT, Gemini, Aira, Seeing AI, smart glasses made for blind users & Cautious use; discloses only presentation needs; re-instructs each session \\
P11 & 45--54 & Blind or low vision (retinitis pigmentosa) & ChatGPT, Gemini, smart speakers, TV assistant & Paid tiers of two assistants; tells them to remember; audits and corrects memory \\
P12 & 65--74 & d/Deaf or hard of hearing; mobility (bilateral hearing loss; back injury) & ChatGPT, Gemini or Copilot & Never disclosed disability; keeps questions general; learned about memory during the interview \\
\bottomrule
\end{tabularx}
\Description{A table with one row per participant, P1 to P12, and five columns: participant ID, age range, self-reported disability with conditions where the participant named them, the assistants they use, and their memory and account situation at the time of the interview. Participants range in age from 25 to 34 through 65 to 74 and include blind and low-vision, mobility, chronic illness, hard of hearing, neurodivergent, and mental health disabilities. Assistants include ChatGPT, Claude, Gemini, Copilot, Siri, Alexa, and smart glasses. The memory column notes, for example, that P1 has access needs written into system instructions, P6 uses ChatGPT without an account by choice, P9 prefers not to be remembered, and P12 learned about memory during the interview.}
\end{table*}

\subsection{Procedure}

The first author conducted all interviews in English over Zoom. Each interview opened with the study's focus stated in plain terms: how participants tell or talk to an assistant about their disability, whether directly by stating it or indirectly by asking for something that only a person with that disability would ask. Interviews were semi-structured and followed a protocol with eight parts: how the participant uses AI assistants; how they handle disclosure in everyday life, when meeting new people, starting a job, or seeing a new clinician; whether and how they have told an assistant about their disability, directly or indirectly; what they expected the assistant to do with that information; experiences of the assistant remembering, forgetting, or resurfacing disability-related information across sessions; unwanted or mistaken assumptions; a direct comparison between disclosing to assistants and to people; and design preferences. In the final part the interviewer introduced design probes (Section~\ref{sec:control}); we report reactions to these as reactions to interviewer-introduced ideas. Participants who had reduced their use were also asked why.

Sessions lasted 44 to 86 minutes (median 62). Access adaptations were used throughout: P1 answered by typing into the meeting chat, and P12 relied on automatic captions, so the interviewer restated and typed key questions. Sessions were audio- and video-recorded with consent confirmed both before and after recording began. Recordings were transcribed automatically; a verbatim transcript was retained for verification and coding. We removed fillers and false starts while preserving the order of turns, participants' phrasing, and every concrete example. Quotes below are lightly edited for readability, and identifying details (names, employers, organizations, locations, and named products that would identify a participant) have been removed or generalized.

\subsection{Data Analysis}

We analyzed the transcripts using reflexive thematic analysis~\cite{braun2006thematic,braun2019reflecting,braun2021onesize}. The first author, who conducted the interviews, read each transcript in full and developed an initial codebook inductively at the level of episodes: a specific instance of disclosing, withholding, being remembered, being misread, correcting, or reacting to a design probe. The first and second authors then independently coded all twelve transcripts with this codebook, the second author working from de-identified versions. The two coders compared their code applications, resolved differences through discussion, and revised code definitions where the comparison showed they were unclear. Codes were then organized using the vocabulary of contextual integrity~\cite{nissenbaum2010privacy}: what type of information was flowing (a diagnosis, a functional need, a symptom, a preference), between which actors (participant, assistant, platform, third parties such as family members or clients), and under what transmission principles (relevance, confidentiality, access control, retention, purpose). Both coders reviewed the resulting codebook and candidate themes against de-identified excerpts and refined the themes through discussion. Consistent with reflexive thematic analysis we did not compute inter-rater agreement; where we report how many participants described an experience, the counts are meant to show the spread of an experience across the sample rather than its prevalence in any population.

\subsection{Positionality and Ethics}

The first author is a disabled scholar with a congenital physical disability who uses a wheelchair and alternative input methods, and who uses LLM assistants daily, including for accessibility. She disclosed this to participants, usually early in the session, and it shaped the interviews. Participants spoke to her as a member of the community, compared assistive setups with her, and in several cases asked her questions in return; the interviews are conversational and sometimes reciprocal, which we believe produced franker accounts of frustration and of ableism than a more distanced stance would have. It also carries a risk of leading: the interviewer occasionally offered her own experiences (for instance, an assistant inserting a stored fact about her typing speed into unrelated emails) as examples when a question was not landing, and we treat the design probes described above as her ideas. Our team includes disabled and non-disabled researchers in accessibility and HCI. We use identity-first and person-first language interchangeably, following our participants' mixed usage~\cite{dunn2015personfirst}, and the singular ``they'' for all participants.

\section{Findings}

We organize the findings around the flow of disability information from a participant to an assistant and onward: what participants disclosed and why (RQ1), who they understood the recipient to be (RQ2), and what memory did with what they said, how they responded, and what control they wanted (RQ3). Each subsection reports one of the six themes from the analysis. Table~\ref{tab:themes} in Appendix~\ref{app:themes} lists the themes with representative codes, the number of participants coded to each, and an illustrative excerpt per code.

\subsection{Theme 1: Disclosure by Need, Not by Name}
\label{sec:need}

Asked whether they had told an assistant about their disability, six of the twelve participants first described something other than a diagnosis. They described an instruction. P10, who is blind and has recently acquired a hearing impairment, does not ``do any inquiries'' about disability at all, but ``I will indicate that I use a screen reader, or that I want them to avoid extraneous punctuation like asterisks.'' P1 has written access needs into an assistant's system instructions: ``when I'm debugging things, the instructions are to give me terminal commands in code boxes so I can just copy and paste.'' P8 tells the assistant why a document needs special handling: ``I'm uploading a PDF, please read it word for word, because I'm blind and this document is not accessible with a screen reader.'' P4 has told ChatGPT that they use a wheelchair, because they ask for wheelchair-accessible outings, but has never named their condition: ``It didn't seem like a requirement \ldots\ I just didn't see it as super necessary. If AI asked, `What disability do you have?' I would've shared. But it didn't come up.''

In the vocabulary of contextual integrity, the \emph{information type} participants chose to send was a functional need scoped to a task, rather than an identity attribute, and for three participants (P4, P6, P10) this scoping was a stated rule rather than an accident of what had come up. P10 stated it as a policy:  ``I only disclose what I need to, meaning I will only explain what I need, how I need information to be presented, which doesn't really tell the AI much about me personally.'' Two participants (P4, P6) explicitly said that they apply the same need-to-know rule to the assistant that they apply to people. P4 noted that their employer of five and a half years has never asked their diagnosis and they have never offered it; with the assistant, ``I think it's pretty much the same reason. Like I said, I just don't find it necessary, or, like, it doesn't super add to the conversation, unless it's extremely relevant. I feel like it's on a need-to-know basis.'' P6, whose disabilities are invisible, has a ``script'' for meeting people that names their conditions and the accommodation they need, and ``just used basically that same approach with ChatGPT,'' scoped by task: ``If I'm asking ChatGPT to help me format a bibliography, I don't have to tell it that I have an anxiety disorder. But if I'm asking it for feedback on something I wrote, then I have to tell it, `Please don't criticize, please be constructive,' because I have an anxiety disorder.'' P6 names the conditions together with the need, by the same logic: in their script the diagnosis is what makes the accommodation intelligible, and because their disabilities are invisible, ``that's why I have to do so much disclosure to let people know I have these access needs.''

Diagnoses were disclosed in two circumstances. The first was when the task itself was medical: P3, who has two rare conditions, uses assistants to interpret lab results and medication options, and P2 told an assistant ``all about'' their muscular dystrophy while building a diet plan. P4, who never named their condition, nonetheless shared symptoms, medications, and timelines during a bout of pneumonia; illness was disclosed where disability was not. The second was when a participant expected the label to carry knowledge the assistant should already have. P7 said ``AI definitely knows that I have a spinal cord injury, because I've explicitly told it: `I have a C4 spinal cord injury. This is my disability.'\,'' P3 explained why they had named their condition rather than listing its consequences: ``I had provided it the context that I have SMA. From my perspective, it should have some knowledge base about SMA, whether or not standing would be possible.'' The diagnosis was offered as a compression of many constraints, on the assumption that a system trained on the internet would decompress it correctly. As Section~\ref{sec:failures} shows, that assumption was not always met.

Two participants (P10, P12) had disclosed little or almost nothing about their disability, and P9 has largely stopped. P12 said ``I don't think I've actually stated in making my inquiry to AI that I have a specific or any disability. I might just ask a general question,'' citing concern about personally identifiable information, but also gave a reason specific to their disability and the medium: ``my engagement with AI is usually via text. So there hasn't really been a need to tell AI I have hearing loss in order for AI and me to communicate. If there was some application where AI and I were interacting vocally, I would probably be more inclined to tell AI I have hearing loss.'' For a hard-of-hearing user, a text interface is already accessible, so disclosure has no instrumental purpose; the need that drives disclosure for blind and physically disabled participants does not arise. P9, who is blind, has found the same need fading as the assistants improved: ``As AI has gone up to where it is now, it's been a lot less necessary to say, `I need you to tell me which one is the garlic salt, because I'm blind.'\,''

Disclosure was not only what participants typed. P8, who creates media about blindness and processes transcripts through ChatGPT, said ``I don't know that I ever disclosed it to ChatGPT directly, I think it picked up on it from all my content,'' and P11 noticed that one assistant ``has latched onto the fact that I have a disability.'' The sender's decision about what to say is only one input to what the recipient comes to know.

One asymmetry between assistants and people ran under all of this. P7, ``very much an open book'' with people, put it precisely: ``AI is a little different, because it doesn't ask questions about it. You just have to tell it. I don't really tell people my disability; I tell AI my disability. I tell people when they ask.'' Because the assistant never asks, disclosure to it is always volunteered, and it is volunteered when a task requires it. P11, whose assistants have memory on and have been told to remember, still states the audience in every prompt: ``never assume it's going to know.'' In answer to RQ1, what participants sent the assistant was usually a need, sometimes a diagnosis offered as shorthand for one, and always volunteered; RQ2 asks whom they understood themselves to be sending it to.

\subsection{Theme 2: One Chat, Two Recipients}
\label{sec:recipients}

When participants compared disclosing to assistants with disclosing to people, they split into two groups with opposite conclusions, and the split turned on who they took the recipient to be. This grouping is our interpretation of which recipient appeared to govern each participant's disclosure decisions.

Six participants (P2, P3, P5, P7, P8, P11) described disclosing to assistants as freer, or less effortful, than disclosing to people. Across these accounts, participants emphasized that the assistant needs the information to help, does not judge, and it costs nothing socially. P2: ``When I'm talking to AI, I'm giving it completely unfiltered, zero-judgment everything, because it's a robot. It's factual. It doesn't have any feelings about it \ldots\ When I talk to humans, it depends on the human \ldots\ What's their situation in my life, are they a caregiver, and I'm their employer, so we have that dynamic?'' P11, who is open about disability with people, nonetheless said ``sometimes it's actually more comfortable being able to communicate with the AI, because you know the AI is not going to judge you like a human''; with AI ``you don't have to worry about the stress of how some humans are when it comes to dealing with your disability.'' P8 compared the assistant with a rideshare driver: ``I'm more interested and more apt to give AI the information that I'm blind or low vision than to give the Uber or Lyft driver that information. Because I've had times where I ordered a Lyft or an Uber, and right after they gave me a driver I'd say, `Okay, listen, I'm blind, I won't see you, please call my name when you arrive,' and they'd cancel.'' P5 inverted the need-to-know rule between audiences: ``I do with AI, because they need to know that in order to help me properly. I disclose all of my different disabilities as I need to. I'm not ashamed of any of them. But I don't disclose any information a person doesn't need.''

These participants were aware that a company sits behind the assistant, and two of them kept concerns about it (P8 about devices that are always listening, P11 about how long data is retained), but none of the six described those concerns as shaping what they told the assistant about disability. P2 had considered the company and set it aside: ``I've thought about it, but I don't really care. I'm not a very private person.'' For this group, the salient recipient was the agent in the conversation, and the norms were those of talking to a discreet, knowledgeable helper.

Three participants (P9, P10, P12) foregrounded the platform, the organization that stores and can reuse what is said, and described disclosing less to assistants than to a person they trusted. P10, who works in assistive technology, framed it as a trade: ``I'm also balancing privacy needs with convenience. I'm careful about how much I disclose, because I'm not very confident that these platforms are careful about protecting my data and my privacy.'' P12 described a history of data breaches and a job with confidentiality rules, and doubted that withholding a name accomplishes much: ``I'm sure the AI is able to determine, based on where the inquiry's coming from, there's some kind of code, they can tell that it's me.'' P9 drew a line at health, ``if I do something medical-related, I wouldn't want that saved,'' and worried about ``people using your words against you \ldots\ I think we've been moving too fast.'' For this group, the discretion of the agent mattered less because the platform was the recipient they were most concerned about. P1 showed how the platform can come into view after the initial disclosure: P1 has written their disability into system instructions because ``it's easier just to have the AI know about it rather than keep explaining,'' yet connected a quick rejection from an internship at the company behind one of their assistants to that openness: ``I do wonder if they looked at my history \ldots\ me talking about all the things I struggle with.''

P6's account adds a third way of experiencing the agent as a recipient that neither literature anticipates, the agent as a judge whose evaluation is itself the harm. P6 uses ChatGPT without an account so that it cannot accumulate a picture of them: ``I don't want ChatGPT to judge me or say anything negative about me, or anything that could be triggering. So I don't want to give it so much information about who I am, where it could then give me an evaluation of who I am.'' An early interaction had settled it: ``I feel like it didn't pass the test. Its answers were not encouraging enough, and it was not supporting my disability pride.'' Where the first group assumed the agent does not judge, P6's experience was that it does, in the same register as people.

Which recipient a participant foregrounded was not simply a matter of temperament. P10 noted that caution is affordable because they have alternatives many blind people lack, a sighted spouse, a paid visual interpreter service, and decades of non-AI strategies: ``I also have available resources that other people don't have.'' P10 also recognized that others may make a different calculation: ``I might as well sacrifice my privacy, because we don't have a lot of privacy anyway.''

Five participants (P6, P9, P10, P11, P12), from both sides of the split, articulated the same transmission principle for the company as recipient: secondary use was more acceptable when it was not tied to identity and served a collective purpose. P10: ``If I can trust that it's not tied to my identity, then I don't care. And actually, there is some benefit to having people with disabilities interact with it, because for at least some of us it will challenge the stereotypic data that's baked into these.'' P9 said information ``should only be for research purposes''; P12, who had disclosed nothing, said they would share more ``if I can contribute to science''; P11 proposed a retention limit of a year. In answer to RQ2, participants applied the same norms they apply to people, but to two different recipients, and which recipient they foregrounded explained whether they disclosed more or less to the assistant than to people.

\subsection{Theme 3: Memory as Relief and Drift}
\label{sec:memory}

Cross-session memory changed the transmission principle of every disclosure: what had been said once, for one purpose, became available to the assistant in all later conversations. Participants experienced this in two ways at once.

The first was relief. Seven participants (P1, P3, P4, P5, P7, P8, P11) welcomed the assistant carrying disability or bodily information forward, because it removed the burden of repeated disclosure; P3, for instance, was pleased that an assistant remembered their body weight when discussing medication doses. P4 recalled the first time: ``it's brought it up in future conversations, like, `Since you mentioned you were in a wheelchair, here are some accessible ideas.' I was like, `That's interesting.'\,'' P7 said ``It feels like it knows me better.'' The clearest statement of memory's value came from a participant whose devices lack it. P5 said of their smart speaker, ``She just cannot remember I'm blind,'' and of the assistant they were about to pay for: ``I don't want to keep reminding ChatGPT that I can't see.'' Two participants pay the cost of repetition on purpose: P6 has no account, and P10 re-instructs each new session, ``if I initiate a new conversation, I have to remind them,'' as the price of not saving.

The second experience was drift. Once stored, disability information reappeared where participants had not sent it, in three patterns. In \emph{over-anchoring}, the assistant applied a stored fact indiscriminately. P2 had told an assistant that caregivers prepare things for them in the morning; it fixed on one detail: ``Now every time it tells me, `Oh, you should have another glass of water this evening before bed,' it also tells me to make sure my caregivers loosen the cap on the bottle every single time.'' It does so when no caregiver is present, and P2 has given up correcting it: ``It's always the last line of every output, so I just scroll past and ignore it at this point.'' P1, planning a trip, found that ``every answer after that was recommending the same wheelchair-taxi company, even if it wasn't relevant to the prompt.'' P3, who receives disability benefits, gets unsolicited warnings: ``sometimes it'll recommend, `Oh, be careful not to affect your Social Security benefit,' when, number one, I'm already aware of that, and number two, I didn't want the insight provided.'' In \emph{misattribution}, the assistant assigned someone else's information to the participant. P11, who researches for clients, found that ``I was asking for something related to a client, and unfortunately it started associating that with me,'' to the point that ``It literally put me as working in organizations that I don't work for.'' In \emph{bleed}, information crossed the boundary between contexts a participant had deliberately separated. P2 keeps a nutrition persona and a mental-health persona in one assistant, but ``when I'm just talking to Gemini randomly and I say, `Hey, my buddy and I are trying to find a place to go to dinner,' it comes back with, `Well, this doesn't fit your diet.' And I'm like, `I'm not asking the diet one!'\,'' 

Two participants (P2, P4) articulated a test of relevance for whether a resurfaced disclosure was appropriate, and P4 added a second test, provenance. P4: ``If it's relevant or related at all, that's fine. If it's totally unrelated, that would probably be off-putting.'' P2 made the same judgment when rejecting per-message control (Section~\ref{sec:control}): relevance cannot be specified in advance. Provenance is a signal that the assistant is drawing on something the participant said, and why. P4 wanted it to say, ``Based on what you said, this reminds me of blah blah, and I'm bringing it up now because of that reason, instead of just randomly bringing up previous conversations.'' Some information should not persist at all, regardless of relevance: P4 did not mind the assistant knowing about skin problems, ``but if I want to ask about, say, feminine health, I might not want them remembering that''; P9 would not want anything medical saved; P3 wanted to switch off one category of stored context, the benefits warnings, while keeping the rest: ``I don't need to get lectures about it every single time.'' P7, by contrast, could not imagine minding: asked about the assistant tailoring advice for a friend to their own needs, they said ``I wouldn't care that it gave me something specialized to me if I'm talking about somebody else.'' For RQ3, memory was welcomed for what it saved participants from repeating and resisted where it let a disclosure travel, and relevance and provenance were the tests participants used to tell the two apart.

\subsection{Theme 4: When the Assistant Does Not Act on What it Knows}
\label{sec:failures}

Drift was one kind of failure. The other, reported by seven participants, was the assistant failing to act on what it knew or could have known: five (P1, P2, P3, P5, P7) had it ignore a stated physical constraint, usually by assuming a non-disabled body; P4 was given an accessibility claim that proved wrong; P9 was misgendered from a name. P1, whose system instructions describe a wheelchair and an eye tracker, found that ``there have been times it completely missed the mark and suggested something that would be unworkable. I think it just isn't able to reason about these things.'' P2, who had told an assistant about their muscular dystrophy while planning a diet: ``there was a time where I was like, `Oh, I don't feel good,' and it said, `Why don't you get up and take a walk?' And I'm like, `You idiot, you know this stuff about me.'\,'' P3, researching a medication after naming their condition, was warned not to stand up too quickly: ``Obviously I have SMA, so I cannot stand up. So, for example, it really should not tell me not to `stand up too quickly' when I'm not able to stand.'' P7's workout suggestions ``downplay'' their injury: ``I'm only able to curl my biceps and things like that, but it might give me stuff that involves using my wrists.'' P5's smart speaker explains the television visually every time.

The failures P6 described were not lapses of memory but of values. Asked, early on, what jobs would suit their access needs, P6 was told, in effect, to toughen up: ``It was basically telling me I should learn to handle feedback better. Instead of asking people not to criticize me, ChatGPT was saying, `Well, feedback is how you learn and grow.'\,'' A writing assistant suggested changing a biographical sentence describing P6 as an author ``with'' multiple sclerosis to an author ``despite having'' multiple sclerosis, ``implying my disability would be an obstacle to what I have achieved. And for me, I'm proud to be disabled.'' P6 did not read these as irrelevant answers: ``It's on topic. It makes sense as an answer \ldots\ I just feel like it's ableist. I've heard that same answer from human beings too, a lot of them have told me I just need to get tougher. So I think ChatGPT is just getting that from its data.''

Participants had theories about why the assistant failed to use what it knew. P3 attributed the warning to liability: ``I think it's so biased to safety that whether or not I can stand doesn't matter, as long as it recommended it, then OpenAI is protected from being liable for something it recommended.'' P6 and P10 attributed ableist answers to training data, ``the stereotypic data that's baked into these.'' P9 explained why they avoid asking assistants disability-related questions at all: ``because it's still human error \ldots\ humans are kind of flawed with the understanding of disabilities and have lower expectations \ldots\ I think humans cannot imagine being what they are not, and so they think of it as, like, hard.'' 

The affective register was mostly annoyance rather than injury. P2 said the lapses are ``more about wasting my time and not being useful than about my disability.'' P5 objected less to the mistake than to the apology: ``Lady, you don't have emotions. Stop giving me emotions and just give me advice. I don't want your apology, I want your advice.'' P6 was the exception, and P6 was also the participant whose disability is most directly implicated in how feedback is delivered. Four participants (P8, P10, P11, P12) reported no disability-related failures of their own. Whether the assistant ignored a stated constraint or reproduced an ableist norm, the effect was the same: a disclosure had been made and not honored, which is what the strategies in the next section respond to.

\subsection{Theme 5: Boundary Work}
\label{sec:boundary}

Between the assistant that forgets and the assistant that over-remembers, participants did considerable work to keep disability information in the contexts where they wanted it. Table~\ref{tab:strategies} summarizes the strategies we observed, in five families.

\begin{table*}[t]
\caption{Strategies participants used to keep disability information in the contexts they intended.}
\label{tab:strategies}
\small
\begin{tabularx}{\textwidth}{@{}L{2.4cm} X L{3.9cm}@{}}
\toprule
\textbf{Family} & \textbf{Strategy} & \textbf{Participants} \\
\midrule
Partitioning & Separate accounts, platforms, or tiers for work, personal, and hobby use & P2 \\
 & Personas, ``gems,'' or reusable workflows that scope what the assistant attends to & P2, P8 \\
Profiling & Writing access needs into system instructions or a profile, or telling the assistant to remember & P1, P2, P11 \\
Auditing & Asking what it knows and correcting it; inspecting memory through developer tools; temporary chats; per-project memory files & P3, P11 \\
Minimizing & Disclosing only presentation needs; phrasing questions generally; withholding identifiers, financial, or medical details & P10, P12, P5, P8, P9, P4 \\
 & Using the assistant without an account, re-disclosing each session & P6 (and P10 by re-instructing) \\
 & A private input channel (braille display, screen and speech off) so others cannot follow & P5 \\
Verifying and repairing & Checking accessibility claims and formatting with trusted sighted people & P4, P8, P9, P10, P11 \\
 & Correcting in the moment; asking the assistant to ask questions first; having one session write the prompt for another & P2, P3, P5, P7, P8, P11 \\
 & Tolerating: scrolling past a recurring irrelevant line & P2 \\
\bottomrule
\end{tabularx}
\Description{A table of strategies participants used to keep disability information in the contexts they intended, in five families with the participants who used each. Partitioning: separate accounts or tiers (P2) and personas or workflows (P2, P8). Profiling: writing access needs into system instructions or telling the assistant to remember (P1, P2, P11). Auditing: asking the assistant what it knows and correcting it, inspecting memory through developer tools, temporary chats (P3, P11). Minimizing: disclosing only presentation needs or withholding identifiers and medical details (P4, P5, P8, P9, P10, P12), using no account (P6), and a private braille input channel (P5). Verifying and repairing: checking accessibility claims with trusted sighted people (P4, P8, P9, P10, P11), correcting in the moment and prompt craft (P2, P3, P5, P7, P8, P11), and tolerating recurring irrelevant lines (P2).}
\end{table*}

\emph{Partitioning} recreates by hand the separation of contexts that memory erodes. P2 uses one assistant for business and another for personal life, a free tier of the first for a hobby, and separate instances for separate clients, rather than trusting one to keep things apart: ``it's easier to have them completely separated than for Claude to say, `Hey, these are pretty much the same, let me connect the dots.'\,'' The reason is asymmetric cost: ``I would rather correct them on the times they don't understand than risk having the overlap when they shouldn't, because untangling the overlap is harder than instructing them on something new.'' \emph{Profiling} is the opposite move, deliberately giving the assistant a durable picture so that it need not be told again; P2 wished for an easier way to ``Build me a profile'' including a schedule. \emph{Auditing} treats memory as an object to be inspected and corrected: P11 periodically asks each assistant what it knows, finds it ``90\% right,'' and fixes the rest with ``correction prompts: `No, this is not me. This was for someone else' \ldots\ So basically, I call that just memory correction.'' \emph{Minimizing} was the strategy of those who foregrounded the company as recipient, and included choosing a channel: P5 connects a braille display to their phone so that ``if I turn my screen and my speech off, the only person who knows what's going on on my phone is me.'' \emph{Verifying and repairing} was the most widespread family: nine participants did one or the other. Five (P4, P8, P9, P10, P11) verified accessibility-critical output before relying on it, most often by routing it past a trusted sighted person (P8 texts a sighted daughter, ``does this look okay?''; a friend checked P9's resume formatting; P10 will ``verify with a human'' anything ``of critical personal importance''), and four (P2, P3, P5, P7) corrected in the moment and moved on.

These strategies were effective and expensive. P2, who worked in software for fifteen years, has decided that teaching the assistant when a caregiver instruction applies is ``more effort than it's worth to me to get AI to dial it in right.'' P3 described a habituated resignation: ``I'm just so used to adapting my environment that I kind of have an apathy toward how they design the app. I just find a way no matter what.'' P3 also named the inequality in who bears the cost. Assistants hide memory controls to keep the interface simple, ``and then that natively impacts people like us, because we need more control with some stuff''; the settings are ``quite buried,'' several clicks deep, so ``it needs to be just one button to manage the memory per chat.'' These strategies restore by hand the context that memory removes, and they were available mainly to the participants best equipped to build them.

\subsection{Theme 6: What Control Would Help}
\label{sec:control}

The interviewer proposed, in varying combinations, three designs: a control on each message for whether the assistant should remember it, personas with separate memories, and memory that expires after a chosen period. Reactions did not sort into for and against. They differed in where participants located the judgment about when disability information should apply: at the moment of capture, in the separation of contexts, in access and retention, in the visibility of stored information, and in the model itself.

Eight participants discussed a \emph{control at the moment of capture} attached to an individual message for whether the assistant should remember it, in some cases combined with per-session scoping or a choice of duration (P1, P2, P3, P7, P8, P9, P11, P12). Four declined it. P2 could ``picture what you're describing, and as a user I would be really frustrated with it, because it couldn't do exactly what I wanted,'' and named the underlying problem: relevance cannot be specified in advance. ``It's when. Can you define when those occur? I can't define that. I know my business one never needs to know about my diet, it doesn't matter.'' P7 said ``I don't think I would ever use it if it was a thing. Because I don't really notice when AI does save stuff to memory, so I wouldn't really notice when it doesn't.'' P8 was ``okay with it the way it is.'' P9 did not want to be remembered at all: ``I don't think I'd want them to remember me in that way. I'm good with me having to do that. It just feels like too much information.'' P1 redirected the question away from the interface: ``maybe it just needs better training overall.'' P12, who learned during the interview that assistants retain memory at all, found such options ``helpful.'' P3 and P11 took up other forms of the idea, described below.

Six participants (P2, P3, P4, P7, P11, P12) discussed \emph{control across contexts}: separate personas or sessions with non-overlapping memory. Reactions differed in kind rather than degree. P2 already partitions contexts by hand (Section~\ref{sec:boundary}) and preferred keeping them separate to letting the assistant connect them. P7 likewise rejected carry-over: ``starting off with a fresh AI every time is good.'' P4, asked about controlling what is shared across sessions, thought such controls would ``probably be overwhelming'' (``Sometimes I don't mind, the less I know, the better''), while still wanting a selective forget control for ``health symptoms, especially if more personal.''

Three participants (P5, P10, P11) located control in \emph{access and retention}: in who could reach stored information and how long it persisted. P5 wanted a single control, over people rather than facts: ``The only thing I want control over is who can access my ChatGPT account. I would rather it remember everything, as long as I'm the only one who can access it.'' P11, offered an individual-message control combined with a choice of duration, took up the duration rather than the toggle, naming it (``I would call that `categorized memory time limits'\,''), endorsing it (``I'm working on this project, and you can forget this after I'm done with it''), and adding a retention limit of a year and a ``Dump my memory'' button. P10, asked whether more control over what is kept would help, said ``if I was able to exert more control, I might be more willing to divulge more personal information,'' and located the strongest form of that control outside the platform: processing ``on device only, so information isn't sent to the cloud.''

Five participants (P3, P4, P10, P11, P12) located \emph{control in visibility}, discussing a visible record of what the assistant stores and of where information used in an answer comes from. All five reacted positively, though each emphasized a different aspect of visibility. P3 wanted memory to be legible and switchable at the point of use, ``just one button to manage the memory per chat,'' rather than several screens deep. P10 said visibility would change their behavior: ``If I could see which portions of my data would be saved, I might be more willing to use it to some extent.'' P11 wanted an easy-to-use interface through which users could actually see stored information; P12 described the ability to see and delete stored memory as ``a safety net, a safeguard.'' P4 wanted the provenance signal described in Section~\ref{sec:memory}, so that the assistant would indicate when a stored fact was shaping an answer and why.

P6, asked a general question about whether more control would help, would find such control helpful only under a prior condition: ``if AI were not so ableist. If I could trust AI with that information.''

Two participants (P3, P7) located \emph{control in the model} itself, preferring that the assistant do more of the contextual reasoning itself rather than leave every decision to the interface; P1's redirection above points the same way. P3 did not ``mind putting a little bit of trust in the AI model.'' P7 argued against disability-specific assistants: ``You're gonna have 4 billion versions of ChatGPT, because everybody's ability is different. What I think should happen is: just make it as good as possible, as accommodated as possible.''

One design idea came from a participant rather than the interviewer. P3, discussing the unsolicited warnings about benefits (Section~\ref{sec:memory}), proposed an expert override for protective boilerplate: ``maybe some people toggle, like, `I am an expert, so I don't need you to,' or `I can opt out of the safety precaution.'\,'' 

In answer to the second part of RQ3, participants located control over who can see stored information (access), which conversations it may enter (scope), how long it persists (retention), whether its use is announced (provenance), and whether protective defaults can be waived (override). 
Nor did any participant spontaneously propose that the assistant ask about disability. P4 said they would have answered if asked, while P10 expressed discomfort with profiling questions.

\section{Discussion}

We interviewed twelve disabled adults about what they tell conversational AI assistants about their disability, how that compares with telling people, and what changes when the assistant remembers. Read through contextual integrity, their accounts show that memory changed not what they disclosed but where a disclosure could travel. We develop that reading first, then turn to what it asks of the systems.

\subsection{Disability Disclosure to Assistants, Read Through Contextual Integrity}

\begin{figure}[!t]
    \centering    \includegraphics[width=0.95\linewidth]{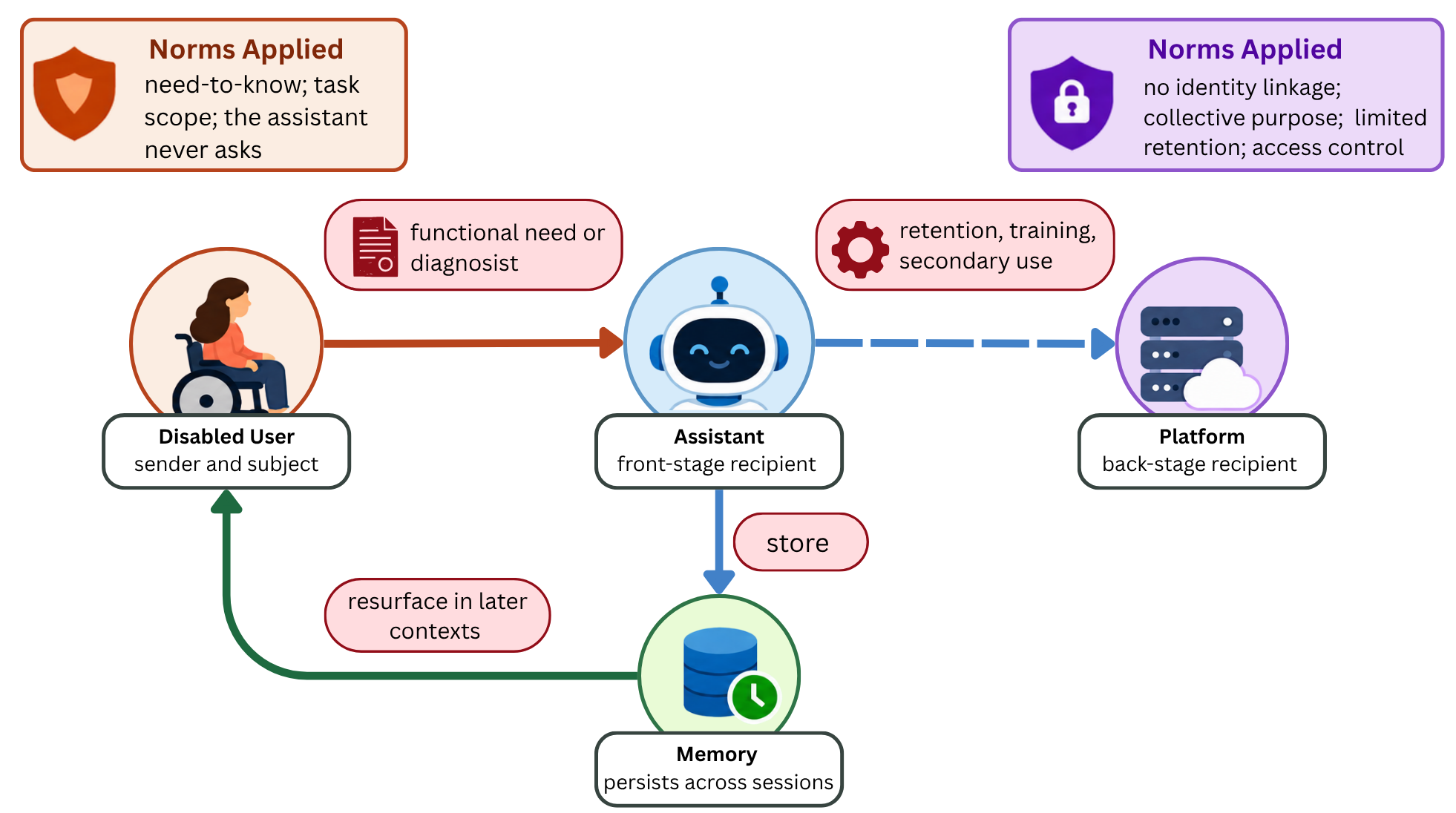}
\caption{Disability information flows in conversational AI as participants described them. A single disclosure is received by two recipients, the assistant in the conversation and the platform behind it, and participants judged it against different norms depending on which recipient they foregrounded. Memory converts an ephemeral, task-scoped flow into a persistent one that can resurface in contexts the sender did not intend.}
\label{fig:flows}
\Description{A diagram with four boxes. On the left, a box labeled Disabled user, sender and subject. An arrow labeled functional need, or diagnosis points right to a box labeled Assistant, front-stage recipient. A dashed arrow labeled retention, training, secondary use points from the Assistant to a box labeled Platform, back-stage recipient. Below the Assistant, a shaded box labeled Memory, persists across sessions, receives an arrow labeled store from the Assistant, and a curved arrow labeled resurface in later contexts returns from Memory to the user. Notes above the user box read: norms applied: need-to-know, task scope, the assistant never asks. Notes above the platform box read: norms applied: no identity linkage, collective purpose, limited retention, access control.}
\end{figure}

Contextual integrity asks of any information flow: what is flowing, from whom to whom, about whom, and under what principle. Our participants' accounts show that each of these parameters behaves unusually when the recipient is a conversational assistant, and that the unusual behavior explains patterns that would otherwise look contradictory. Figure~\ref{fig:flows} summarizes the picture.

\paragraph{The sender chooses the information type, and it is usually a need.} Six participants described telling an assistant an instruction rather than a diagnosis, and the five who named a condition did so for medical tasks or because they expected the label to carry knowledge the assistant should already have. This is data minimization performed by the user rather than the system, continuous with disclosure compelled by access rather than chosen as an expression of self~\cite{samuels2003body,lingsom2008invisible}, and it complicates a common assumption in work on personalization, that knowing a user's disability lets a system adapt: what our participants needed was not an assistant that knew their disability but one that followed an instruction. Naming a diagnosis was a bargain in which the participant compressed many constraints into a label and trusted the model to decompress it, and the model sometimes did not (Section~\ref{sec:failures}). Prior work has documented ableist associations in language models~\cite{hutchinson2020social,gadiraju2023offensive,phutane2025cold}; our data show one route by which those associations could reach disclosure itself. We say ``could'' because no participant reported abandoning a named diagnosis after such a failure: P3 and P7 kept naming their conditions and corrected the assistant when it erred. The clearer effect was on what people chose to share at all. P6 withheld a profile after early answers that were ``not supporting my disability pride,'' and P9 avoids the assistant for disability-related questions because it inherits human ``lower expectations.'' 

\paragraph{The trigger for disclosure is reversed.}
Disclosure to people is largely reactive: a visible aid, a question, a moment of communicative breakdown prompts it~\cite{shinohara2011shadow,evans2019trial}. Disclosure to an assistant is proactive, because, as P7 put it, ``it doesn't ask.'' This reversal puts the burden of anticipating relevance on the disabled user, and it explains why memory was attractive to seven participants despite its costs. Memory is the only mechanism by which the assistant ever knows without being told again, which for users who type slowly (P1, P7), or who are tired of re-explaining (P5), is an accessibility feature rather than a convenience. Designers who respond to the privacy problems of memory by weakening it should notice who pays for that choice.

\paragraph{There are two recipients, and participants' norms were consistent once we knew which one they meant.}
The divergence in our data, between participants who found disclosing to assistants freer than disclosing to people and participants who disclosed less, was not a divergence in norms. It was a divergence in the recipient to which a shared norm was applied.
Those who foregrounded the agent treated it as a discreet, knowledgeable helper who needs the facts and does not judge, the same reasoning that leads people to disclose more to computers they believe are non-judging~\cite{lucas2014computer,ho2018chatbot} and that makes the human-like interface a driver of disclosure~\cite{zhang2024fairgame}. Those who foregrounded the platform applied the norms that people apply to institutions: no linkage to identity, purpose limitation, limited retention. The two sets of norms differ in kind, and the difference is itself a finding. Assistant-side norms were situated and practice-based: participants could say which task warranted which disclosure, in what words, and how a resurfaced fact should be handled, because they had concrete, repeated experience interacting with assistants. Platform-side norms were general and institutional, because participants had no comparable relationship with the platform; what it does with their data was inferred from experiences such as breaches and from expectations about organizations in general. 
This doubled recipient is, we think, the single most useful concept for reading survey results in which disabled respondents report using AI readily for sensitive tasks and yet, once a company would receive the data, strongly prefer a human instead~\cite{afb2026innovation,afb2026quagmire}. Both are true, of different recipients. P6's account adds a third way of experiencing the agent as a recipient that neither literature anticipates, the agent as a judge whose evaluation is itself the harm. P10's account also reminds us, with Sannon and Forte~\cite{sannon2022privacy}, that foregrounding the platform is a stance one must be able to afford: caution was easiest for the participant with the most alternatives to AI.

\paragraph{Memory changes the transmission principle from ephemeral to persistent, and participants have folk tests for when that is appropriate.}
Contextual integrity predicts that a flow which was appropriate under one transmission principle becomes problematic when the principle changes~\cite{nissenbaum2010privacy}. Memory does exactly this: a functional need shared to complete one task becomes a standing attribute applied to every task. The three patterns of drift we observed, over-anchoring, misattribution, and bleed, are the disability-specific face of what memory benchmarks now call cross-domain leakage and memory-induced sycophancy~\cite{pulipaka2026persistbench,zhang2024ghost}.
What our participants add is a pair of appropriateness tests that a system could operationalize. A resurfaced disclosure is acceptable when it is relevant to the current task, and it is more acceptable when its provenance is announced. Neither test is a rule about categories of information; both are rules about the relation between a stored flow and the present context. This is the sense in which contextual integrity, applied to memory, is not a checklist~\cite{shvartzshnaider2025position}. P2's remark that ``I can't define'' in advance when a fact should apply is a precise statement of why per-message controls fail: the norm lives in the future context, not at the moment of capture.

\paragraph{Boundary work is the cost of a missing abstraction.}
The strategies in Table~\ref{tab:strategies} are, almost without exception, manual reconstructions of the contexts that memory flattens: separate accounts recreate separate recipients, personas recreate scoped flows, audits recreate visibility into what the recipient holds, and verification with a trusted sighted person recreates the accountable human recipient that the assistant is not~\cite{alharbi2024misfitting}. That the most technically expert participants did the most of this work, and said that ``people like us'' need more control, echoes long-standing findings that the burden of access is distributed unevenly and is often invisible to those who design the systems that create it~\cite{bennett2020care,hofmann2020living}.

\subsection{Implications for the Design of Conversational Assistants}
\label{sec:implications}

The findings suggest design directions at three layers of an assistant: the interaction that captures a disclosure, the model that acts on it, and the platform that holds it. The directions for scoping, point-of-use control, and acting on functional statements rest on several participants' accounts; those for provenance, access control, retention limits, and an expert override rest on one or two, and we present them as proposals to test rather than requirements. Several run counter to the per-message ``remember this'' controls that assistants have begun to offer, which four of the eight participants who discussed them declined (Section~\ref{sec:control}).

\subsubsection{Interaction: Capturing Disclosure}

\paragraph{Treat a functional statement as the unit of disclosure.}
Participants disclosed needs rather than names (Section~\ref{sec:need}), so the assistant's job is to recognize a functional statement as an access need, confirm that it has registered it, and honor it for as long as it applies. This removes the diagnosis from the loop, which is what participants wanted, and makes the disclosure portable across sessions and products. When a diagnosis is offered anyway, the assistant should treat it as a compressed set of constraints to be unpacked cautiously, with explicit uncertainty about variation within a condition, rather than as a label that licenses assumptions. Two things follow. The assistant should not ask about disability, since no participant wanted a profiling question and the need-to-know rule participants apply to people (Section~\ref{sec:need}) applies here too. And it should not be specialized by disability; the target is one general assistant that reliably acts on stated needs.

\paragraph{Lower the cost of volunteering.}
Because the assistant never asks, every access need has to be volunteered, and volunteering is expensive for people who type slowly or whom voice modes interrupt (Section~\ref{sec:memory}). 
Capturing access needs once and reusing them, and building voice interaction that waits for a slow or interrupted speaker, lower that cost more than any memory control would.

\paragraph{Surface inferences as disclosures.}
Assistants infer disability from content and behavior (Section~\ref{sec:need}), and an inferred attribute is a disclosure the sender never made. When a system infers a disability or an access need, it should show the inference and let the person confirm, correct, or discard it, so that what the recipient holds stays traceable to what the sender chose to say. Two participants noticed such inferences, and analyses of real memory stores suggest the pattern is common~\cite{dash2026algorithmic}.

\subsubsection{Model: Acting on What It Has Been Told}

\paragraph{Check outputs against stated constraints.}
The commonest failure was the assistant recommending something the user had already said they cannot do (Section~\ref{sec:failures}). A consistency check that asks, before a recommendation is delivered, whether it conflicts with any stated constraint is cheap, and one participant had already watched a model audit its own output for accessibility issues; the same mechanism extends to bodily constraints.

\paragraph{Treat feedback style as an access need, and evaluate for ableist defaults.}
Some failures were not lapses of memory but of values: advice to toughen up, or edits that recast disability as an obstacle (Section~\ref{sec:failures}). An instruction about how feedback should be delivered is, for some users, as much an access need as a screen-reader instruction, and should be honored as durably. Because these defaults come from training data that reflects human ableism~\cite{gadiraju2023offensive,phutane2025cold,sharma2024sycophancy}, evaluation should include disabled people, whose interaction with the model, as one participant noted, is itself a corrective to the stereotypic data it learned from.

\paragraph{Mark accessibility claims as unverified, or cite them.}
Participants already route accessibility-critical output past a trusted person or verify it themselves (Section~\ref{sec:boundary}), a practice documented elsewhere~\cite{alharbi2024misfitting}. An assistant that flagged accessibility claims as unverified, or cited their source, would meet that practice rather than leave users to discover errors on arrival.

\paragraph{Let users waive protective boilerplate for conditions they name.}
Warnings that contradict a stated constraint read as not listening (Section~\ref{sec:failures}). A user-declared expertise setting, scoped to a named condition and reversible, would let the assistant drop repeated warnings for that condition while keeping them elsewhere. Tentative, since it rests on one participant's proposal, and the scoping is the whole design: an unscoped override would suppress warnings a user needs.

\subsubsection{Platform: Holding the Disclosure}

\paragraph{Make both recipients legible at the point of disclosure.}
Participants decided what to say according to which recipient they had in mind (Section~\ref{sec:recipients}), so retention, use in training, and who can see conversations should be visible in the conversation, where the decision is made, rather than in terms of service. A prominent no-retention mode belongs there, and cautious participants said they would disclose more under terms they could see~\cite{sharma2025beforemom}.

\paragraph{Scope memories to the context they came from.}
Memory changes a disclosure's transmission principle from task-scoped to global, and drift follows (Section~\ref{sec:memory}). Stored facts should carry the context in which they were disclosed and a default scope, so that a diagnosis shared while asking about medication is treated differently from an access need written into a profile. Participants already build scoping by hand with accounts, personas, and per-project files (Section~\ref{sec:boundary}), and persona-based agents have been proposed to reduce interaction friction for people with motor disabilities~\cite{taheri2023virtual}; the system should offer it as a default the user can widen, not a partition the user must remember to select.

\paragraph{Put memory controls where the conversation is.}
Participants who wanted control wanted it at the point of use, not several screens deep (Section~\ref{sec:control}). One control per conversation, showing what will be kept and letting the user change it, is the form they described, and because these controls are used by screen-reader and alternative-input users, their accessibility is part of the privacy design.

\paragraph{Announce provenance, and make correction one gesture.}
When a stored fact shapes an answer, the assistant should say so briefly and offer a single action to say \emph{not here} or \emph{not anymore}. This turns each resurfacing into a lightweight, in-context norm elicitation, which is where participants located the norm (Section~\ref{sec:memory}): relevance is judged in the moment, not specified in advance. Tentative, since one participant asked for the signal itself, but it is the mechanism that would have resolved the misattribution and bleed others described without a settings screen.

\paragraph{Treat accounts as shared, and set retention by category and time.}
Accounts and devices are shared with family and used on behalf of clients (Section~\ref{sec:recipients}), so assistants should support distinct people under one account and keep one person's access needs from being applied to another. Retention defaults by category (medical and intimate health information excluded by default) and by time match the institutional norms participants applied to the platform (Section~\ref{sec:recipients}). Both are tentative, resting on a few accounts, and both would let cautious users disclose what a task needs.

\paragraph{Support the no-account path.}
Some participants pay for privacy with repetition (Section~\ref{sec:boundary}). Session-scoped or device-local profiles that hold access needs without persisting them to a server, and on-device processing where feasible, would give them the relief that memory gives everyone else without asking them to trust the platform as a recipient.

\subsection{Limitations and Future Work}

This is an interview study of twelve people, and its claims are about patterns of reasoning, not prevalence. Participants were, as a group, experienced with technology, and the boundary work we describe is theirs, not a typical user's. The two-recipient pattern in particular should be read with care, since two of the three participants who foregrounded the platform were also among the lightest users, so what looks like a difference in which recipient people reason about may partly reflect how much experience they have had with these systems. Nine participants were White, and most had had their disabilities since birth or early childhood; people with acquired disabilities, people with intellectual disabilities, and Deaf signers were absent or thinly represented, and each group may reason about disclosure differently. All participants were in the United States, where assistants' data practices and disability law differ from those in other locations. 

The interviews relied on recall. Participants described disclosures and resurfacings from memory, and the drift we report is what they noticed, not what occurred; inferences the assistant made silently, of the kind recent analyses of memory stores document, would be invisible to this method. The interviewer's own disability produced candid conversations, and in a few places the example a participant discussed followed one the interviewer offered; the design probes were hers, and reactions to them should be read as reactions rather than as independent design ideas. We coded the transcripts. Finally, the products we studied change monthly; memory features that were new to some participants may behave differently by the time this paper is read. 

Future work should observe disclosure and resurfacing in situ through logs or diary methods rather than recall; deploy scoped memory and provenance signals with disabled users to test whether the relevance and provenance tests we identified can be operationalized; and co-design with the communities we did not reach.

\section{Conclusion}

Disabled people have always managed disability information carefully, deciding what to tell, to whom, and in what words. Conversational AI assistants receive those disclosures under conditions that the disability literature did not anticipate: the recipient never asks, is two parties at once, and remembers. Our twelve participants responded by disclosing needs rather than names, by applying consistent norms to whichever recipient they had in view, and by doing considerable work to keep what they said in the context where they said it. What they asked for was not an assistant that knows less but one that lets a disclosure travel only as far as it should: that acts reliably on a stated need, keeps a fact within the context it came from, says when a memory is in play, lets the right people and no others converse under a profile, and trusts them as experts on their own bodies. Contextual integrity gave us a vocabulary for these requests; the participants supplied the norms. Designing assistants to respect them is, we have argued, an accessibility problem as much as a privacy one.

\section*{Use of Generative AI}

Generative AI tools were used to support the development of this manuscript, including language editing, organization, and refinement of author-written text. The authors reviewed and verified all AI-assisted text. The authors take full responsibility for the accuracy, interpretation, and integrity of the manuscript.

\begin{acks}
    We are grateful to the twelve participants who gave their time and spoke so candidly about their experiences. We thank the disability organizations, listservs, and online communities that circulated our recruitment call, including the National Federation of the Blind's research participant request program, the SMA Support System group on Facebook, and the Association of Late-Deafened Adults and its Facebook discussion group. We also thank everyone who helped us reach participants, and Ezra Awumey for his thoughtful feedback on an earlier draft.
\end{acks}

\bibliographystyle{ACM-Reference-Format}
\bibliography{reference}


\onecolumn

\appendix
\section{Themes, Codes, and Illustrative Excerpts}
\label{app:themes}

\begin{table}[h]
\caption{The six themes with three representative codes each and one illustrative excerpt per code. The full codebook has 73 codes; $n$ is the number of participants with at least one excerpt assigned to the code.}
\label{tab:themes}
\small
\renewcommand{\arraystretch}{1.15}
\begin{tabularx}{\textwidth}{@{}>{\raggedright\arraybackslash}p{3.0cm} >{\raggedright\arraybackslash}p{4.6cm} X@{}}
\toprule
\textbf{Theme} & \textbf{Code ($n$)} & \textbf{Illustrative excerpt} \\
\midrule
Disclosure by Need, Not by Name (Section~\ref{sec:need})
 & Functional need or presentation instruction (7) & P1: ``I told it about sometimes being fatigued: it should not give me instructions that involve clicking around or switching between windows.'' \\
 & Diagnosis, condition, or status named (5) & P6: ``I say, `I identify as a person with a disability.' I tell ChatGPT that I have MS \ldots\ and I explain the access needs.'' \\
 & The assistant does not ask; disclosure is volunteered (3) & P11: ``I have found with these tools: never assume it's going to know.'' \\
\midrule
One Chat, Two Recipients (Section~\ref{sec:recipients})
 & Assistant as non-judging helper; freer disclosure than to people (6) & P5: ``It's best to share as much as you can without getting creepy, because that's the only way they can really help you.'' \\
 & Platform or company as recipient (5) & P10: ``The profit motive and the competitive nature of these companies mitigates against privacy.'' \\
 & Purpose limitation and collective benefit (5) & P11: ``I don't think they should retain the data for more than a year at a time.'' \\
\midrule
Memory as Relief and Drift (Section~\ref{sec:memory})
 & Welcomed resurfacing; relief from repetition (7) & P8: ``With Meta Ray-Bans, that seems to have stuck at this point. I don't have to say it every time.'' \\
 & Over-anchoring or over-correction (3) & P1: ``Every answer after that was recommending the same wheelchair-taxi company, even if it wasn't relevant to the prompt.'' \\
 & Cross-context bleed (2) & P2: ``It comes back with, `Well, this doesn't fit your diet.' And I'm like, `I'm not asking the diet one!'\,'' \\
\midrule
When the Assistant Does Not Act on What It Knows (Section~\ref{sec:failures})
 & Ignores a known physical constraint (5) & P2: ``When it knows my situation and still tells me to, you know, get up and do something, like, take my dog for a run.'' \\
 & Wrong accessibility claims or hallucination (3) & P4: ``Sometimes it says something is accessible, and then I go verify, and it's not really accessible.'' \\
 & Attributed cause: training data, human ableism, lower expectations (3) & P9: ``Humans are kind of flawed with the understanding of disabilities and have lower expectations.'' \\
\midrule
Boundary Work (Section~\ref{sec:boundary})
 & Profiling: system instructions, profile, tell it to remember (3) & P2: ``There's a top-level knowledge base, always remember this about me, and I fill that in with, `This is my situation.'\,'' \\
 & Auditing: ask what it knows, correct, inspect (2) & P3: ``If you open the developer tools and look at the payload, you can literally see what they remember about you.'' \\
 & Verifying with trusted people or sources (5) & P9: ``I've never fully trusted it. I use it as a companion.'' \\
\midrule
What Control Would Help (Section~\ref{sec:control})
 & Wants point-of-use control or visibility (4) & P3: ``It is quite buried in the user interface. You have to click on two buttons, then a third button.'' \\
 & Rejects per-message or fine-grained control (5) & P7: ``I think it would actually make it worse. I think starting off with a fresh AI every time is good.'' \\
 & Prefers the model to reason contextually (2) & P3: ``The state-of-the-art models are quite intelligent now. I don't mind putting a little bit of trust in the AI model.'' \\
\bottomrule
\end{tabularx}
\Description{A three-column table with the columns Theme, Code with the number of participants coded to it, and Illustrative excerpt. It has six theme groups, each with three representative codes and one short verbatim participant quotation per code, labeled with the participant ID. Theme 1, Disclosure by Need, Not by Name: functional need or presentation instruction, seven participants; diagnosis, condition, or status named, five; the assistant does not ask and disclosure is volunteered, three. Theme 2, One Chat, Two Recipients: assistant as non-judging helper with freer disclosure than to people, six; platform or company as recipient, five; purpose limitation and collective benefit, five. Theme 3, Memory as Relief and Drift: welcomed resurfacing and relief from repetition, seven; over-anchoring or over-correction, three; cross-context bleed, two. Theme 4, When the Assistant Does Not Act on What It Knows: ignores a known physical constraint, five; wrong accessibility claims or hallucination, three; attributed cause in training data, human ableism, or lower expectations, three. Theme 5, Boundary Work: profiling through system instructions, a profile, or telling the assistant to remember, three; auditing by asking what it knows, correcting it, or inspecting stored data, two; verifying with trusted people or sources, five. Theme 6, What Control Would Help: wants point-of-use control or visibility, four; rejects per-message or fine-grained control, five; prefers the model to reason contextually, two.}
\end{table}

\end{document}